\documentclass[preprint,showpacs,preprintnumbers,amsmath,amssymb,nofootinbib]{revtex4}
\usepackage{graphicx}
\usepackage{dcolumn}
\usepackage{bm}
\usepackage{color}
\usepackage{epsfig}
\def\PL #1 #2 #3 {{\it Phys. Lett.} {\bf#1} (#3) #2}
\def\NP #1 #2 #3 {{\it Nucl. Phys.} {\bf#1} (#3) #2}
\def\ZP #1 #2 #3 {{\it Z. Phys.} {\bf#1} (#3) #2}
\def\PRL #1 #2 #3 {{\it Phys. Rev. Lett.} {\bf #1} (#3) #2}
\def\PR #1 #2 #3 {{\it Phys. Rev.} {\bf#1} (#3) #2}
\def\MPL #1 #2 #3 {{\it Mod. Phys. Lett.} {\bf#1} (#3) #2}
\def\RMP #1 #2 #3 {{\it Rev.~Mod. Phys.} {\bf#1} (#3) #2}

\newcommand{\be}{\begin{equation}}
\newcommand{\ee}{\end{equation}}
\newcommand{\ba}{\begin{eqnarray}}
\newcommand{\ea}{\end{eqnarray}}

\newcommand{\e}{\epsilon}

\newcommand{\beqn}{\begin{eqnarray}}
\newcommand{\eeqn}{\end{eqnarray}}
\newcommand{\beqns}{\begin{eqnarray*}}
\newcommand{\eeqns}{\end{eqnarray*}}

\newcommand{\beq}{\begin{equation}}
\newcommand{\eeq}{\end{equation}}
\newcommand{\beqa}{\begin{eqnarray}}
\newcommand{\eeqa}{\end{eqnarray}}

\newcommand{\nn}{\nonumber \\}

\newcommand{\ep}{\epsilon}

\begin{document}
\preprint{FERMILAB-PUB-08-177-T}
\preprint{UH-511-1128-08}
\title{ 
Masses, fermions and generalized $D$-dimensional unitarity
}

\author{R.~Keith~Ellis}
\email{ellis@fnal.gov}
\affiliation{ Fermilab, Batavia, IL 60510, USA }
\author{Walter~T.~Giele}
\email{giele@fnal.gov}
\affiliation{ Fermilab, Batavia, IL 60510, USA }
\author{Zoltan Kunszt}
\email{kunszt@itp.phys.ethz.ch}
\affiliation{Institute for Theoretical Physics, ETH, CH-8093 Z\"urich, Switzerland}
\author{Kirill~Melnikov} 
\email{kirill@phys.hawaii.edu}
\affiliation{Department of Physics and Astronomy,
          University of Hawaii,\\ 2505 Correa Rd. Honolulu, HI 96822}  

\date{\today}
\begin{abstract}
We extend the generalized $D$-dimensional unitarity method 
for numerical evaluation of one-loop amplitudes by 
incorporating  massive particles. 
The issues related to extending the spinor algebra to higher dimensions,
treatment of external self-energy diagrams and mass renormalization 
are discussed within the context of the
$D$-dimensional unitarity method.
To validate our approach,  we calculate in QCD the one-loop 
scattering amplitudes of a massive quark pair with up to three
additional gluons for arbitrary spin states of the external quarks 
and gluons.
\end{abstract}
\maketitle

\section{Introduction}

Good understanding of  background and signal processes
will be necessary to interpret data from the Large Hadron 
Collider (LHC) and observe physics beyond the Standard 
Model.
In particular, large multiplicity
final states are of interest~\cite{Bern:2008ef}.
Reliable predictions for such processes require 
computations of next-to-leading order (NLO) QCD corrections.
Traditional methods for NLO calculations have difficulties 
in dealing with processes of such complexity; as a result, 
many new approaches to one-loop computations 
have been suggested in  recent years~\cite{Bern:2008ef}.

Among those approaches,  generalized unitarity 
stands out~~\cite{Bern:1994zx,Bern:1995db,Bern:1996je,Bern:1996ja,Britto:2004nc,Britto:2004nj,Bern:2005hs,
Bern:2005ji,Bern:2005cq,Berger:2006ci,Britto:2005ha,Britto:2006sj,Mastrolia:2006ki,Anastasiou:2006gt,
Anastasiou:2006jv,Britto:2006fc,Britto:2007tt,Britto:2008vq,Britto:2008sw,Forde:2007mi,
Kilgore:2007qr,Ellis:2007br,Giele:2008ve}. The key feature of this method is that 
it allows calculation  of one-loop scattering amplitudes directly from 
tree amplitudes leading to a computational algorithm
of polynomial complexity~\cite{Bern:2008ef:EGK}.
The efficiency of generalized unitarity 
for NLO calculations  for processes with high multiplicity 
final states has been explicitly demonstrated 
in Refs.~\cite{Berger:2008sj,Giele:2008bc}.

Until recently, generalized 
unitarity was mostly used to compute the cut-constructible 
parts~\cite{Bern:1994cg} of scattering amplitudes, while calculations 
of the
rational parts proved to be challenging. 
In Refs.~\cite{Bern:2005cq,Berger:2006ci} 
the four-dimensional boot-strap 
method was developed to evaluate the rational part.
Another approach developed
to generate the rational part
uses generalized 
$D$-dimensional unitarity
~\cite{Anastasiou:2006gt,Anastasiou:2006jv}.

In a recent paper~\cite{Giele:2008ve}, we 
extended the method
of Refs.~\cite{Ossola:2006us,Ellis:2007br}
in such  a way that {\it both} cut-constructible and 
rational parts are  obtained within a single formalism 
using integer-dimensional on-shell cuts.
This method leads to a computational algorithm of polynomial 
complexity, as shown in Ref.~\cite{Giele:2008bc}.

Up to now, generalized unitarity has been mainly studied in the 
context of multi-gluon scattering amplitudes which simplifies 
the problem significantly. In the general case, one has to deal with 
two additional issues -- different types of particles that 
participate in the scattering process and the fact that massive 
particles can be involved.  It is necessary to address these issues 
before generalized unitarity becomes a practical tool for NLO 
calculations of phenomenological interest.
The goal of this paper is to do exactly that and  
extend the applicability of
generalized $D$-dimensional unitarity by considering 
one-loop amplitudes involving gluons and massive quarks.
The computational method developed
in Ref.~\cite{Giele:2008ve} can handle both extensions easily.

Dealing with particles of different flavors requires more sophisticated
bookkeeping, but is otherwise straightforward. However, the presence of massive 
particles introduces new conceptual issues. An  
obvious consequence of having virtual particles with non-zero masses contributing 
to one-loop scattering amplitudes is that 
in addition to quadruple, triple and double cuts, 
we also have to deal with  single-particle cuts. Such 
an extension is straightforward; the necessary details have already been 
given in Ref.~\cite{Ellis:2007br}.
A more interesting  consequence of massive particles present 
in the scattering 
process is that generalized unitarity applied to certain double-
and single-particle cuts becomes more subtle. This is closely
related to external wave function renormalization constants 
which originate from Feynman diagrams with self-energy insertions 
on external lines\footnote{Similar problems appear due to diagrams that 
can be interpreted as one-loop expectation values of quantum fields.}.  
We will show that this complication can be circumvented without encumbering
the formalism. 

To validate the method, we focus on the calculation of one-loop amplitudes 
with a massive quark anti-quark pair and up to three gluons. 
These one-loop amplitudes have been calculated using more
traditional methods. The one-loop corrections to 
$t\bar{t}\ +$ 2 and $t \bar t +$ 3 partons scattering 
have been first calculated in Ref.~\cite{Nason:1987xz,Beenakker:1988bq} and  
 Ref.~\cite{Dittmaier:2008jg}, respectively.

The outline of the paper is as follows. In Section II we discuss the
modification of the $D$-dimensional generalized unitarity method required
to include massive fermions.
Section III describes the subtleties that arise when massive 
particles are involved in the one-loop scattering amplitude.
In Section IV we present numerical results for the one-loop amplitudes 
$0 \mapsto t \bar t\ +$ 2 gluons and $ 0 \mapsto t \bar t\ +$ 3 gluons.
The conclusions and outlook are given in Section V.

\section{One-Loop Amplitudes and Dimensionality of Space-Time}

One-loop calculations in quantum field theory are divergent 
and require regularization at intermediate stages of the calculations.
The conventional choice is dimensional regularization where 
 momenta and 
polarization vectors of unobserved virtual particles 
are continued to arbitrary dimensions~\cite{'t Hooft:1971fh,'t Hooft:1972fi}.
By keeping the momenta and polarization vectors of all 
observable external particles in four dimensions, one
can define the one-loop helicity 
amplitudes to be used in NLO parton-level generators~\cite{Giele:1991vf}. 
Once the dependence
of a one-loop amplitude on the dimensionality of space-time is 
established,  the dimensionality $D$ can be interpolated 
to the non-integer value $D=4-2\e$. The divergences of one-loop 
amplitudes are regularized  by the  parameter $\e$.

While the analytical implementation of the dimensional regularization
procedure is well-established (see for example Ref.~\cite{collins}), 
a numerical implementation needs more
consideration. In Ref.~\cite{Giele:2008ve}
we developed numerical implementation of 
dimensional regularization. To explain the method, we note that 
any $N$-particle one-loop scattering amplitude $A_N^{[1]}$ 
can be written as a linear combination of 
the so-called master integrals. The coefficients of such an expansion 
depend on $D$; this dependence can be made 
explicit by choosing  the appropriate basis of master integrals.
After dimensional continuation, the final expression in the  
four-dimensional helicity (FDH) scheme~\cite{Bern:1991aq,Bern:2002zk}
is given by~\cite{Giele:2008ve}
\beqa\label{MasterDecomp}
&& {A}_N^{[1]}=
\sum_{[i_1|i_5]}\e\times e_{i_1i_2i_3i_4i_5}\ 
I^{(D+2)}_{i_1i_2i_3i_4i_5} \nn
&&+\sum_{[i_1|i_4]} \left(
d_{i_1i_2i_3i_4}\ I^{(D)}_{i_1i_2i_3i_4}
+\e\times \hat{d}_{i_1i_2i_3i_4}\ I^{(D+2)}_{i_1i_2i_3i_4}
-\e(1-\e)\times \hat{\hat{d}}_{i_1i_2i_3i_4}\ I^{(D+4)}_{i_1i_2i_3i_4}
\right)\nonumber \\
&&+\sum_{[i_1|i_3]} \left(c_{i_1i_2i_3}\ 
I^{(D)}_{i_1i_2i_3}+\e\times \hat{c}_{i_1i_2i_3}\ 
I^{(D+2)}_{i_1i_2i_3} \right)
+\sum_{[i_1|i_2]} \left( b_{i_1i_2}\ 
I^{(D)}_{i_1i_2}+\e\times \hat{b}_{i_1i_2}\ 
I^{(D+2)}_{i_1i_2}\right)\nn &&
+\sum_{[i_1|i_1]} a_{i_1}\ I_{i_1}^{(D)}\, 
\eeqa
where we introduced the short-hand notation $[i_1|i_n] = 1\leq
i_1<i_2<\cdots<i_n \leq N$.  The master integrals 
in Eq.~(\ref{MasterDecomp}) are defined as
\beq
I_{i_1\cdots i_m}^{(D)}=\int\frac{d^Dl}{i\pi^{D/2}}\frac{1}{d_{i_1}\cdots d_{i_m}}\ ,
\eeq
with $d_i=d_i(l)=(l+p_1+\cdots p_i)^2-m_i^2$.
The coefficients
$b_{i_1i_2}$, $\hat{b}_{i_1i_2}$, $c_{i_1i_2i_3}$,
$\hat{c}_{i_1i_2i_3}$, $d_{i_1i_2i_3i_4}$,
$\hat{d}_{i_1i_2i_3i_4}$, $\hat{\hat{d}}_{i_1i_2i_3i_4}$, and
$e_{i_1i_2i_3i_4i_5}$ are independent of the dimensionality.

We can compute these dimension-independent coefficients numerically, 
within  the method of $D$-dimensional generalized unitarity.
To accomplish  this,  a parametric integration 
method~\cite{Ellis:2007br,Giele:2008ve},
based on the ideas  developed in Ref.~\cite{Ossola:2006us}, is employed.
The key point is to extend the dimensionality of the loop momentum
to an {\it integer} $D$-dimensional value. 
For one-loop calculations, 
an extension to five dimensions is sufficient \cite{'t Hooft:1971fh,Giele:2008ve}.
However, 
care has to be taken with the dimensional
dependence of the spins of the internal particles. The dimensional
regularization scheme allows us to choose the  dimensionality
for internal degrees of freedom of virtual particles
$D_s$ to be equal or larger than the embedded loop-momentum dimensionality.
By choosing the parametric form of the integrand in integer $(D_s,D)$ dimensions
we can determine the dimension-independent coefficients through
partial fractioning of the integrand. The partial
fractioning factorizes the calculation of the one-loop amplitude
into tree amplitudes~\cite{Giele:2008ve}. 
This factorization property is equivalent to the factorization obtained
in generalized unitarity methods.
The four-dimensional helicity scheme defines the parametric continuation as
$D_s\rightarrow 4$, $D\rightarrow 4-2\e$ with the constraint $D_s\geq D$,
giving the final result of Eq.(\ref{MasterDecomp}). 

We consider now the one-loop scattering amplitude involving a massive
quark pair in addition to the gluons: $ 0 \mapsto t \bar t\ +\ N$ gluons.
The $D_s$-dependence of the amplitude is linear.
This means that we need to compute the integrand 
for two different values of $D_s$ so that we can
separate the $D_s$-dependent and $D_s$-independent parts~\cite{Giele:2008ve}.
Because we need well-defined states for fermions  when taking the internal
fermion propagator on-shell, we must choose the space-time 
dimensionality to be even, i.e.  $D_s=4$, $D_s=6$ and $D_s=8$.

The on-shell internal gluonic polarization states in six and eight dimensions
with the momentum vector in five dimensions are straightforward generalizations
of the choices made in Refs.~\cite{Giele:2008ve,Giele:2008bc}.
The construction of $D_s$-dimensional on-shell fermionic lines requires an extension of the
four-dimensional Clifford algebra. We need to explicitly construct the 
$D_s$ Dirac matrices $\Gamma^{\mu}$ and the $2^{D_s/2-1}$ 
spin polarization states $u_j^{(s)}(l,m)$ that satisfy the Dirac equation 
\be
\sum_{j=1}^{2^{D_s/2}}(\sum_{\mu=0}^{D-1} l_\mu \Gamma_{ij}^\mu - m\times\delta_{ij} ) u_j^{(s)}(l,m) = 0, 
\ee
and the completeness relation 
\be
\sum \limits_{s = 1}^{2^{(D_s/2-1)}} u_i^{(s)}(l,m) \bar u_j^{(s)}(l,m)
= \sum_{\mu=0}^{D-1} l_\mu \Gamma_{ij}^\mu + m\times\delta_{ij}\ , 
\label{eq.comp}
\ee
where the on-shell condition for a fermion with the 
mass $m$ and momentum $l$ reads $l^2=m^2$.  
To construct the explicit higher dimensional Dirac matrices we
follow the recursive definition given in Ref.~\cite{collins}.
The $8 \times 8$ six-dimensional Dirac matrices are defined in terms of 
the $4\times 4$ four-dimensional Dirac matrices $\{\gamma^0,\gamma^1,\gamma^2,\gamma^3,\gamma^5\}$
\be
\Gamma^0 = 
\left (\begin{array}{cc}
\gamma^0 & 0 \\
0 & \gamma^0 
\end{array}
\right ),\;\;\;
\Gamma^{i=1,2,3} =
\left (
\begin{array}{cc}
\gamma^i & 0 \\
0 & \gamma^i 
\end{array}
\right ),\;\;\; 
\Gamma^{4} =
\left (
\begin{array}{cc}
0  & \gamma^5 \\
-\gamma^5 & 0
\end{array}
\right ),\;\;\; 
\Gamma^{5} =
\left (
\begin{array}{cc}
0  & i \gamma^5 \\
i \gamma^5 & 0
\end{array}
\right )\ .
\label{eqd1}
\ee
It is readily checked that these matrices satisfy the standard 
anti-commutation relation 
\be
\Gamma^{\mu} \Gamma^{\nu} + \Gamma^{\nu} \Gamma^{\mu} = 2 g^{\mu \nu},\;\;\;\;
\mu,\nu=0,\ldots,5.
\label{eq2}
\ee
The $16 \times 16$ eight-dimensional Dirac matrices are constructed 
in a similar manner from the
six-dimensional Dirac matrices.
The $D_s$-dimensional Dirac matrices are given 
for a particular representation of the 
Dirac algebra. Other representations can be obtained by unitary transformations.
To construct a set of $2^{D_s/2-1}$ spinors satisfying the Dirac equation we
generalize the procedures used in the four-dimensional case. We define the spinors
\be
u^{(s)}(l,m) = \frac{(l_\mu \Gamma^\mu + m)}{\sqrt{l_0 + m}}\eta_{D_s}^{(s)},
\;\;\;\; s=1,\ldots,2^{D_s/2-1}\ .
\ee
For $D_s=4$ we choose
\beqa
\eta_4^{(1)}=\left (\begin{array}{c} 1 \\ 0 \\ 0 \\ 0 \\ \end{array}\right),\;\;\;\
\eta_4^{(2)}=\left (\begin{array}{c} 0 \\ 1 \\ 0 \\ 0 \\ \end{array}\right)\ ,
\eeqa
and construct recursively the $D_s=6$ eight-component basis spinors
\be
\eta_6^{(1)} = \left ( \begin{array}{c} \eta_4^{(1)} \\ 0 \end{array} \right ),\;\;\;\;
\eta_6^{(2)} = \left ( \begin{array}{c} \eta_4^{(2)} \\ 0 \end{array} \right ),\;\;\;\;
\eta_6^{(3)} = \left ( \begin{array}{c} 0 \\ \eta_4^{(1)} \end{array} \right ),\;\;\;\;
\eta_6^{(4)} = \left ( \begin{array}{c} 0 \\ \eta_4^{(2)} \end{array} \right )\ .
\ee
The eight spinors for $D_s=8$ are obtained using the obvious generalization.  
It is easy to see that the spinors constructed in this way do 
indeed satisfy the Dirac equation.

To check the completeness relation, we need the Dirac-conjugate spinor 
$\bar u$.
One subtlety associated with the fact that we have to  deal with 
complex, rather than real, on-shell  momenta is that in order to satisfy 
the completeness relation Eq.~(\ref{eq.comp}), 
we have to {\it define} the conjugate spinor as 
\be
\bar u^{(s)}(l,m) = \bar \eta_{D_s}^{(s)} 
\frac{(l_\mu \Gamma^\mu + m)}{\sqrt{l_0 + m}}.
\label{eq.ncc}
\ee
Note that the loop momentum is {\it not} complex conjugated. It is then 
straightforward to check that the completeness relation 
Eq.(\ref{eq.comp}) is satisfied.

\section{Massive particles and the unitarity cuts}

To determine the dimension-independent master integral
coefficients in Eq.~(\ref{MasterDecomp}) we use
the $D$-dimensional generalized unitarity method of Ref.~\cite{Giele:2008ve}. 
To this end, we parameterize the integrand of the one-loop amplitude
\beq\label{parametric}
{\cal A}_N^{[1]}(l)=
\sum_{[i_1|i_5]} 
\frac{\overline{e}^{(D_s)}_{i_1i_2i_3i_4i_5}(l)}{d_{i_1}d_{i_2}d_{i_3}d_{i_4}d_{i_5}}
+\sum_{[i_1|i_4]} 
\frac{\overline{d}^{(D_s)}_{i_1i_2i_3i_4}(l)}{d_{i_1}d_{i_2}d_{i_3}d_{i_4}}
+\sum_{[i_1|i_3]} 
\frac{\overline{c}^{(D_s)}_{i_1i_2i_3}(l)}{d_{i_1}d_{i_2}d_{i_3}}
+\sum_{[i_1|i_2]} \frac{\overline{b}^{(D_s)}_{i_1i_2}(l)}{d_{i_1}d_{i_2}}
+\sum_{[i_1|i_1]} \frac{\overline{a}^{(D_s)}_{i_1}(l)}{d_{i_1}}\,.
\eeq
The left hand side of the equation is completely specified by the Feynman rules. 
The parametric form
on the right hand side of the equation depends on a set of coefficients. To determine
the coefficients for a given phase space point
we use partial fractioning. This isolates the individual pole 
structures, thereby dividing the sets of linear equations to be solved
into smaller subsets. More importantly, the partial fractioning sets 
groups of internal
lines on-shell. This organizes the left hand side 
of the equation into products
of gauge invariant tree amplitudes, thereby
removing the necessity to compute 
individual Feynman diagrams to evaluate ${\cal A}_N^{[1]}(l)$
for a given loop momentum.

This procedure can  readily  be applied in a situation when massive particles 
are involved in the scattering process.
The presence of massive particles creates more types of master integrals 
or, equivalently, 
more different denominator structures in Eq.~(\ref{parametric}).
Furthermore, the single-cut (or tadpole) contributions to one-loop amplitudes have to be calculated so 
that the tadpole coefficient in Eq.~(\ref{MasterDecomp}) can be determined.
These issues complicate the bookkeeping, but do not add conceptual difficulties.

\begin{figure}[t!]
\includegraphics[angle=270,scale=0.75]{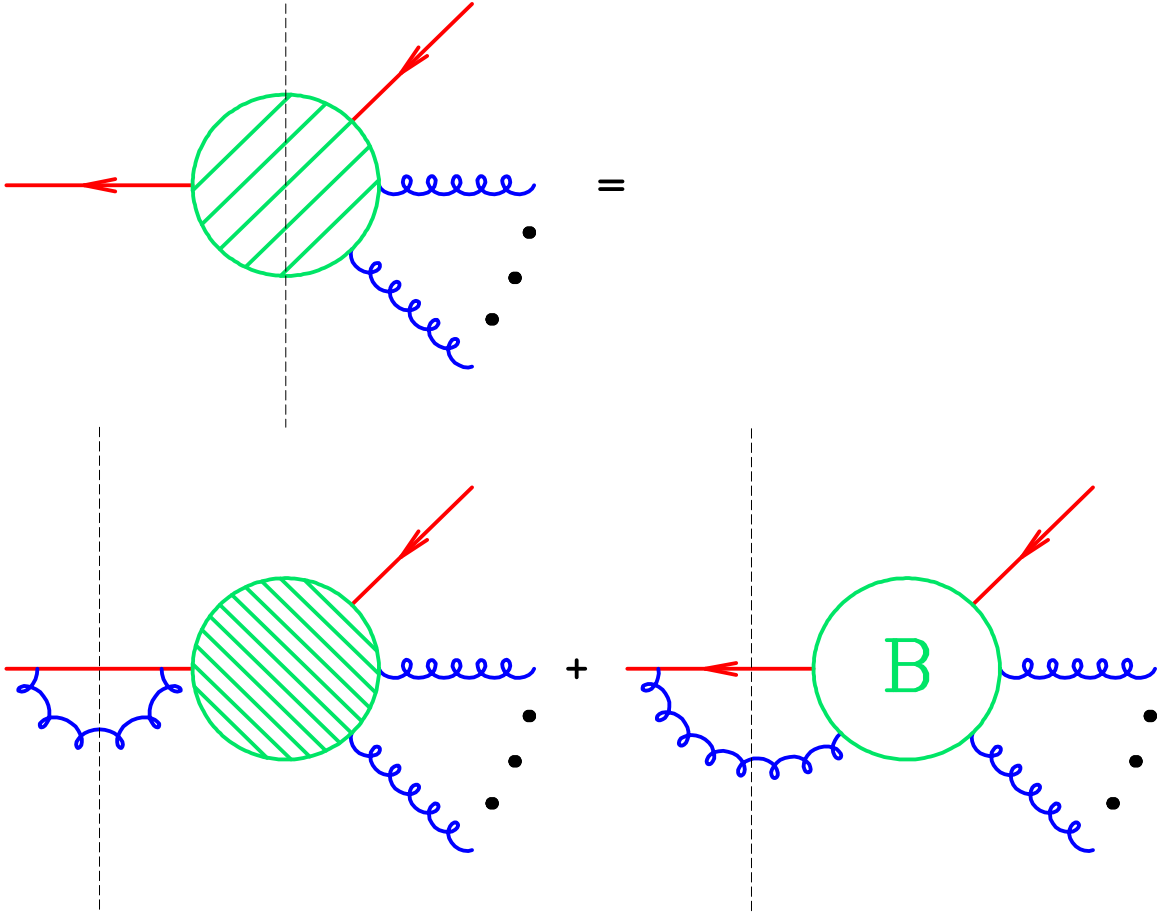}
\caption{The external self-energy cut of a general $t\bar{t}\ +\ N$ gluon loop amplitude
splits in the external self-energy contribution and the remaining higher point contributions.
The different shadings of the blobs represent different content.}
\label{fig:ttfig}
\end{figure}

However, a new conceptual issue does appear
when dealing with  the double cuts shown 
in Fig.~\ref{fig:ttfig}. Note that such cuts need only 
be considered for external massive states, since, if the external on-shell 
line carries 
a light-like momentum, the cut  in Fig.~\ref{fig:ttfig} is 
set to zero 
in dimensional regularization. For massive particles these cuts do give 
non-vanishing contributions.
The subtlety arising when such cuts are 
considered is related to a conflict between 
generalized unitarity and self-energy insertions on the external lines
\footnote{While we discuss the one-loop amplitude  $t \bar t\ + N$ gluons, 
other processes with  massive external lines can be treated in the same way.}.

To see this, we study  the contribution from a particular two-particle
cut shown in Fig.~\ref{fig:ttfig}.
The only outgoing external line to the left 
of the cut is the top quark and all other external particles are to the 
right of the cut.  The residue of the one-loop amplitude for such a cut
can be schematically written as 
\be
{\rm Res} \left [ {\cal A}^{[1]}(t,g_1,\ldots,g_n,\bar{t}) \right ]  
\sim \sum_{\mbox{states}} {\cal A}^{[0]}(t,g^*,\bar{t}^*)\times {\cal A}^{[0]}(t^*,g^*,g_1,\ldots,g_n,\bar{t})\ , 
\label{eq.2.1}
\ee
where $t^*$ and $g^*$ denote the top quark and gluon cut lines respectively and the sum is over
the intermediate states of the on-shell top quark and gluon particles of the two cut lines.
The factorized on-shell tree amplitudes are given by
${\cal A}^{[0]}(t,g^*,\bar{t}^*)$ and  
${\cal A}^{[0]}(t^*,g^*,g_1,\ldots,g_n,\bar{t})$.
However, the latter amplitude is not defined. Separating  the
cut self-energy contribution as indicated in Fig.~\ref{fig:ttfig} gives for
the tree amplitude
\be
{\cal A}^{[0]}(t^*,g^*,g_1,\ldots,g_n,\bar{t}) = \frac{R(t^*,g^*,g_1,\ldots,g_n,\bar{t})}{(p_{t*} + p_{g^*})^2 - m_t^2}
+ B(t^*,g^*,g_1,\ldots,g_n,\bar{t}).
\label{eq3.2}
\ee
Momentum conservation forces 
the invariant mass of $t^* + g^*$ to be equal to the top quark  mass squared,
$(p_{g^*} + p_{t^*})^2 = m_t^2$ making the 
one-quark reducible part of the amplitude singular.

The singular contribution corresponds to the 
self-energy correction to the external top quark line. 
When one-loop scattering amplitudes are calculated using conventional
Feynman diagrams, these type of one-particle reducible diagrams 
are discarded; their effects on the scattering process are 
accommodated later through the external particle 
wave function renormalization constants.  We would like to   
follow this approach in conjunction with the generalized unitarity technique,
but then  care has to be taken with the gauge invariance. 

Suppose we subtract the first term  in Eq.~(\ref{eq3.2}) 
from the tree amplitude; in recursive calculations this 
can be done by truncating the recursive steps. It is then easy to 
see that the remaining part of the amplitude $B$, the second term
in Eq.~(\ref{eq3.2}), is no longer gauge invariant. Indeed, 
the discarded part of the amplitude is 
related to the self-energy correction on the external top quark line; 
such self-energy 
corrections produce  on-shell mass and wave-function renormalization 
factors.
While the mass renormalization constant, $Z_m$, is independent of the gauge-fixing parameter,
the on-shell wave-function renormalization factor, $Z_2$, is {\it not}.
For this reason we have to ensure that the gauge used in calculating the
second term in Eq.~(\ref{eq3.2})
and the gauge  used in the calculation of the wave-function  
renormalization factor $Z_2$ are the same.  
Since the  wave-function renormalization factors 
are most easily computed in the Feynman gauge, we use this gauge to calculate
the residue in Eq.~(\ref{eq.2.1}). This means that the sum over gluon
particle states for the cut in Fig.~\ref{fig:ttfig} includes 
non-physical states $e_s^{\mu}$ such that
\beq
\sum_{s=1}^{D_s} e_s^{\mu}e_s^{\nu}=-g^{\mu\nu}\ .
\eeq
Note that since the offending cuts never involve gluon self-couplings,
ghosts do not need to be considered.

Finally, we note that,  for the most part, 
the coefficients in Eq.~(\ref{parametric}) are computed 
using  the
standard sums over physical states of the 
on-shell particles associated with  cut lines. 
However, for a limited set of pole terms
which contain the external self-energy contributions we need
the procedure described in this Section. We emphasize 
that the conflict between unitarity and self-energy corrections 
to external massive lines  is generic; 
it appears  in any calculation of one-loop scattering 
amplitudes provided that  massive internal or external particles 
are present.

\section{Scattering amplitudes at one-loop}
\begin{figure}[t!]
\includegraphics[angle=270,scale=0.75]{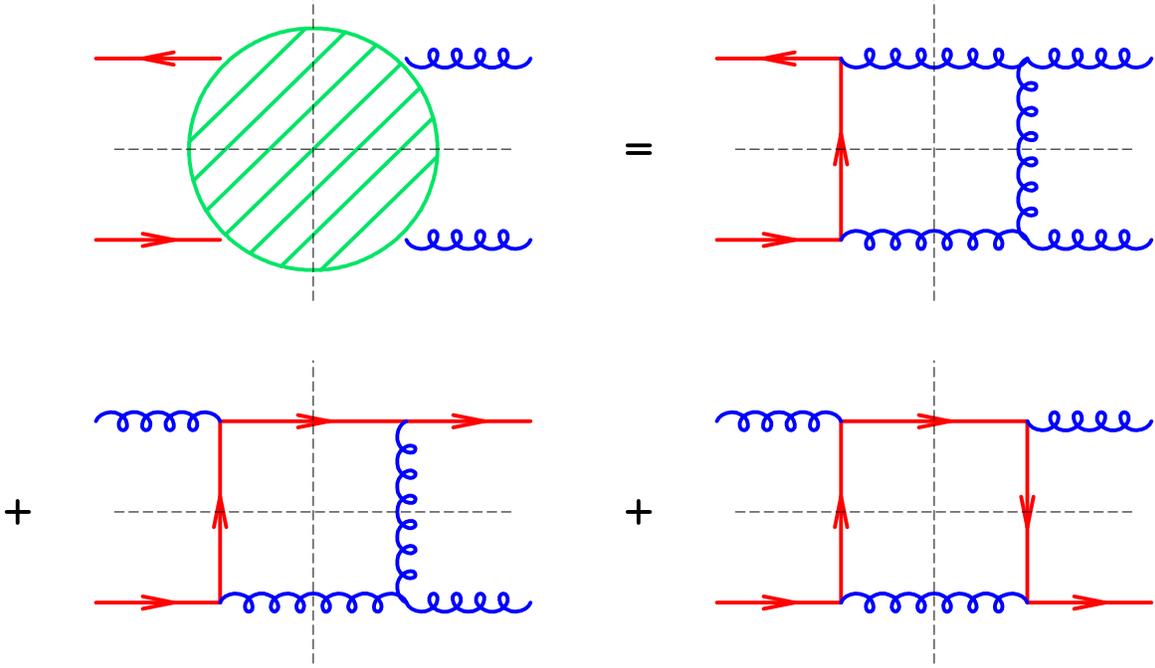}
\caption{The quadruple cut of the $t\bar t$ + 2 gluon amplitude decomposes 
into 
3 gauge invariant contributions, each with its own 4-point master integral. The
first box integral 
contributes to the primitive amplitude $A_L(1_{\bar t}, 2_t, 3,4)$,
the second to  $A_L(1_{\bar t}, 3,2_t,4)$ and the third to 
$A_L(1_{\bar t}, 3,4,2_t)$.}
\label{fig:primitive}
\end{figure}

To implement the generalized unitarity method in a numerical algorithm,
we decompose the $0\mapsto t\bar t\, +\, 2,3$ gluon amplitude into so-called
primitive amplitudes~\cite{Bern:1994fz}\footnote{We adopt the conventions and
normalizations of Ref.~\cite{Bern:1994fz} to define the primitive amplitudes.}. 
Within the context of $D$-dimensional
unitarity the primitive amplitudes play a special role. Each primitive amplitude has 
unique
unitarity cuts, i.e. the flavor of the cut lines is uniquely defined. This is
shown in Fig.~\ref{fig:primitive} for the example of the
quadruple cuts applied to the  $0\mapsto t\bar t\, +\, 2$ gluon 
amplitude.  This quadruple cut  decomposes into three distinct
gauge invariant cuts, each with its own master integral. Each of the
three individual cuts contributes to one of the three primitive 
amplitudes $A_L(1_{\bar t},2_t,g_1,g_2)$, $A_L(1_{\bar t},g_1,2_t,g_2)$ and $A_L(1_{\bar t},g_1,g_2,2_t)$. 

The method described in this paper is amenable to straightforward 
numerical implementation. To evaluate a primitive amplitude we consider
all pole terms in the partial fractioning of Eq.~(\ref{parametric}). 
Double pole terms that correspond to massless 
two-point functions for light-like incoming momenta and single pole terms that
correspond to massless 
tadpoles  are discarded since the corresponding master integrals vanish in dimensional 
regularization. 
The tree amplitudes for each cut are computed using 
Berends-Giele recurrence 
relations~\cite{Berends:1987me}. 
Because single particle cuts contribute,  we need to evaluate the
high multiplicity tree amplitudes $\bar t t\,+\, \bar t t\,+\,2,3$ gluons

Before discussing numerical results for  one-loop
$0\mapsto t\bar t\, +\, 2,3$ gluon amplitudes, 
we remind the reader that, when massive particles are involved, 
additional renormalization constants are required to arrive at physical predictions.
In particular, for massive quarks, on-shell mass and wave function 
renormalization constants are necessary\footnote{Note that the {\it on-shell} 
wave-function renormalization constant contains both ultraviolet 
and infrared divergences. Both show up as poles in $\e$.}. 
For consistency, we need those 
constants in FDH scheme.  As described above, the wave function renormalization 
constant needs to be computed in the Feynman gauge. The bare quark mass $m_0$ 
and the bare quark field $\psi_0$ are renormalized multiplicatively
\be
m_0 = Z_m m, \;\;\;\;\;\; \psi_0 = \sqrt{Z_2} \psi.
\ee
We find $(D=4-2 \epsilon)$
\beqn \label{eqz}
 Z_m = Z_2 &=& 1 - C_F g_s^{2} c_{\Gamma} 
\left ( \frac{\mu^2}{m^2} \right )^\epsilon
\left ( \frac{D_s+2}{2 \epsilon} + \frac{D_s+6}{2} \right ) +O(g_s^4,\epsilon)
\nonumber \\
&\to& 1 - C_F g_s^{2} c_{\Gamma} 
\left ( \frac{\mu^2}{m^2} \right )^{\ep}
\left (  \frac{3}{\epsilon} + 5 \right )+O(g_s^4,\epsilon)\ ,
\eeqn
where $g_s$ is the bare strong coupling constant, $c_\Gamma$ 
is the normalization factor, 
\be
c_\Gamma = \frac{\Gamma(1+\ep) \Gamma(1-\ep)^2}{(4\pi)^{2-\ep} 
\Gamma(1-2\ep)}\ ,
\label{eq4.1a}
\ee
$C_F = (N_c^2-1)/(2N_c)$ is the color factor and 
$\mu$ is the scale which is introduced 
in dimensional regularization to maintain proper dimensionality 
of the coupling constant.  Note that in the last step in Eq.~(\ref{eqz})
we used $D_s = 4$, as is required in the FDH scheme.

We now present the results of the numerical evaluation of 
one-loop $0\mapsto t\bar t\, +\, 2,3$ gluon 
scattering amplitudes in QCD.  We do not include diagrams with 
closed fermion loops.
In addition, external wave function renormalization constants 
and the coupling constant renormalization factors are not included.
However, we do include the mass counter-term 
diagrams which are necessary to obtain a result which is invariant 
under gauge transformations of the external gluons.  For presentation purposes, it is convenient to 
normalize one-loop primitive amplitudes to tree-graph primitive amplitudes
\beqa
{\cal A}_{L}^{[1]}(1_{\bar t},3,\ldots,j-1,2_{t},j,\ldots,n)
&=& c_\Gamma  \left( \frac{a_{L}^{(j)}}{\ep^2} + \frac{b_{L}^{(j)}}{\ep}+ c_{L}^{(j)} \right)
{\cal A}_{L}^{[0]}(1_{\bar t},3,\ldots,j-1,2_{t},j,\ldots,n) \nn
{\cal A}_{L}^{[1]}(1_{\bar t},3,\ldots,n,2_{t})
&=& c_\Gamma  \left( \frac{a_{L}^{(n)}}{\ep^2} + \frac{b_{L}^{(n)}}{\ep}+ c_{L}^{(n)} \right)
{\cal A}_{L}^{[0]}(1_{\bar t},3,\ldots,n,2_{t})\ .
\label{eq4.1}
\eeqa
The coefficients $a_{L}^{(j)}$ and $b_{L}^{(j)}$ parameterize divergences
of the one-loop scattering amplitude. They can 
be extracted from Ref.~\cite{Catani:2000ef}
\beqa\label{catani}
\frac{a_L^{(j)}}{\ep^2} + \frac{b_L^{(j)}}{\ep} &=& \frac{1}{2\ep} 
-S_{\bar{t}g}(p_2,p_j,\mu)-\sum \limits_{i=j}^{n-1} S_{gg}(p_i,p_{i+1},\mu)-S_{gt}(p_n,p_1,\mu) \nn  
\frac{a_L^{(n)}}{\ep^2} + \frac{b_L^{(n)}}{\ep} &=& \frac{1}{2\ep} -S_{\bar{t}t}(p_2,p_1,\mu)\ .
\eeqa

The functions $S_{f_i,f_{i+1}}= S_{f_{i+1},f_i}$ 
depend on the flavor of particles $f_{i}$ ,  their momenta $p_i$ 
and the scale $\mu$. They read
\beqa
S_{t\bar t}&=& \frac{1}{\e\beta}\left(\frac{1}{2}\ln\left(\frac{1-\beta}{1+\beta}\right)
+ i\pi\Theta(d_{t \bar t})\right), \nn \\
S_{tg} &=& S_{\bar tg}=\frac{1}{2\ep^2} 
+ \frac{1}{\e}\left(\frac{1}{2}\ln\left(\frac{m_t^2\mu^2}{d_{tg}^2}\right) + i\pi\Theta(d_{tg})\right), \nn \\
S_{g_1g_2} &=& \frac{1}{\ep^2} 
+ \frac{1}{\ep}\left(\ln\left(\frac{\mu^2}{|d_{g_1g_2}|}\right) + i\pi\Theta(d_{g_1g_2})\right),
\ea
where $d_{i,j} = 2\,p_i \cdot p_{j}$ and 
$\displaystyle \beta = \sqrt{1 -\frac{4m_t^4}{d_{t \bar t}^2}}$.

Finally,  we need to
define the spin states of the gluons and top-quarks. 
For the gluons we use the conventional definition of the helicity vectors
\begin{eqnarray}
p_{\mu}&=&E \big( 1,  \sin\theta \cos\phi,\sin\theta \sin\phi,\cos\theta\big ) \nonumber \\
\varepsilon_{\mu}^{\pm}(p)&=&\frac{1}{\sqrt{2}}\big(0, \cos\theta\cos\phi\mp i \sin\phi
, \cos\theta\sin\phi\pm i \cos\phi,-\sin\theta \big)\ .
\end{eqnarray} 
For the massive on-shell quarks 
($p = (E, p_x, p_y,p_z) ,\;\;p^2=m^2$)
we use the spinors 
\be
u_{+}(p) = \sqrt{E+m} 
\left ( \begin{array}{c}
1 \\ 0 \\ \displaystyle{\frac{p_z}{E+m}} \\ \displaystyle{\frac{p_x+i\,p_y}{E+m}} 
\end{array}
\right),\
u_{-}(p) = \sqrt{E+m} 
\left ( \begin{array}{c}
0 \\ 1 \\ \displaystyle{\frac{p_x-i\,p_y}{E+m}} \\ \displaystyle{\frac{-p_z}{E+m}} 
\end{array}
\right )
\ee
\be
v_{+}(p) = \sqrt{E+m} 
\left ( \begin{array}{c}
\displaystyle{\frac{p_z}{E+m}} \\ \displaystyle{\frac{p_x+i\,p_y}{E+m}} \\ 1 \\ 0  
\end{array}
\right),\
v_{-}(p) = \sqrt{E+m} 
\left ( \begin{array}{c}
\displaystyle{\frac{p_x-i\,p_y}{E+m}} \\ \displaystyle{\frac{-p_z}{E+m}} \\ 0 \\ 1
\end{array}
\right ).
\ee

The  numerical results reported below 
are obtained in conventional double precision
using a {\sf FORTRAN} 77 program. 
The evaluation time does not depend 
on the helicities of the external particles but it 
does depend on the specific primitive amplitude. It takes less time to 
evaluate  primitive amplitudes where quarks are adjacent, than to evaluate
primitive amplitudes where quarks are separated by gluons.
The reason for this is that it is computationally more expensive 
to have more quarks involved in the evaluation of the primitive 
tree amplitudes.
That is, the more quark propagators there are, 
 the longer the evaluation time. 

For evaluating the master integrals
we use the {\sf QCDLoop} program  developed in Ref.~\cite{Ellis:2007qk}.
We have verified that our calculations
correctly reproduce the divergent parts of primitive amplitudes,  
given in Eq.~(\ref{catani}).
For all primitive amplitudes we have 
checked the gauge invariance by substituting a polarization vector 
of one of external gluons by its momentum.
In addition we performed a Feynman diagram-by-diagram check on
the results of the calculation.

\subsection{Scattering amplitudes with two quarks and two gluons}

\begin{tiny}
\begin{table}[th!]
\begin{center}
\begin{tabular}{|c|c|c|c|}
\hline\hline
Amplitude & tree & $c^{\rm cut}$ &  $c$ \\ \hline\hline
$+_{\bar t},+_{t},+_{3},+_{4}$ 
& 0.0026595\;i
& 5.859738 +11.04762\; i
& 43.74436+11.04762\;i  \\ \hline
$+_{\bar t},+_{t},-_{3},+_{4}$ 
& -0.127261\;i
& 18.49057-2.63910\;i & 18.49058
-2.63910\;i \\ \hline
$+_{\bar t},-_{t},+_{3},-_{4}$
& 1.259555\;i
& 20.69972-0.144581 \;i&
   20.52783 -0.14458\;i \\ \hline 
$+_{\bar t},-_{t},-_{3},+_{4}$ 
&  -0.4198517
& 22.16788 -3.40322\;i  & 22.68356 -3.40322\; i 
 \\ \hline \hline

$+_{\bar t},+_{3},+_{t},+_{4}$ 
& 
 -0.0035643 \;i
& -0.26303343   &
-0.26303305  \\ \hline
$+_{\bar t},-_{3},+_{t},+_{4}$ 
& 0.170558 \;i
& 15.2990066  & 15.2990071 \\ \hline
$+_{\bar t},+_{3},-_{t},-_{4}$
&  -1.688090 \;i
&  20.8261462 &  20.8261462 \\ \hline
$+_{\bar t},-_{3},-_{t},+_{4}$ 
& 0.56269666 \;i
& 22.0890527 
 &   22.0890523
 \\ \hline \hline

$+_{\bar t},+_{3},+_{4},+_{t}$ 
& 0.000905\; i
& -26.24047 + 40.67377\; i  & -123.4438 + 40.67377 \;i
  \\ \hline
$+_{\bar t},-_{3},+_{4},+_{t}$ 
& -0.043298 \; i
& 20.00357  -1.69128\;i  
 & 20.00357  -1.69128\;i \\ \hline
$+_{\bar t},+_{3},-_{4},-_{t}$
& 0.4285350\;i 
&21.83688-4.01097\; i 
   & 21.33165 -4.01097 \;i \\ \hline
$+_{\bar t},-_{3},+_{4},-_{t}$ 
& -0.142845\; i
& 19.48471 -1.85887\;i
 & 21.00038 -1.85887 \; i 
 \\ \hline \hline
\end{tabular}
\caption{\label{2g} 
The $0 \mapsto \bar t t\,+\, 2$ gluons results for 
the primitive tree amplitude and finite parts of the three one-loop
primitive amplitudes for various helicities of gluons and top-quarks. 
Both the cut-constructible and total finite one-loop terms are given.}
\vspace{-0.1cm}
\end{center}
\end{table}
\end{tiny}

In Table~\ref{2g} we present the results for the finite parts 
of some of the primitive amplitudes 
$A_{L}(1_{\bar t},2_{t},g_3,g_4)$,
$A_{L}(1_{\bar t},g_3,2_{t},g_4)$ and 
$A_{L}(1_{\bar t},g_3,g_4,2_{t})$ in the FDH scheme. 
The numerical results are obtained for the scale choice $\mu = E$.  
We take the mass of the top quark to be $m_t = 1.75$ and choose 
the following kinematic point
 ($p = (E,p_x,p_y,p_z)$)
\ba
&&p_1=E\left ( 1,0,0,\beta  \right),\;\;\;p_2= E\left ( 1,0,0,-\beta  \right), \nonumber \\
&&p_3=E\left ( -1,\sin\theta,0,\cos\theta \right),\;\;\
p_4=E\left (-1,-\sin\theta,0,-\cos\theta \right ).
\ea
with $E=10$, $\beta = \sqrt{1-m_t^2/E^2}$ and $\theta=\pi/3$. Note 
that  all  the external momenta are taken to be outgoing.
 
On a standard  
Pentium 2.33~GHz processor, it takes about 7.5, 11 and 12 ms 
to evaluate the primitive amplitudes 
$A_L(1_{\bar t}, 2_t, 3,4)$, $A_L(1_{\bar t}, 3,2_t,4)$
and $A_L(1_{\bar t}, 3,4,2_t)$ respectively.  
Approximately  half of that time is spent on the calculation of 
rational parts.  We note that the calculation of the cut-constructible 
part of color-ordered four-gluon amplitude~\cite{Ellis:2007br}
takes about $1~{\rm ms}$. The  difference in CPU time between
$\bar t t\,+\, 2$ gluons and  the four gluon  amplitude is not dramatic.
The time difference is the result of several factors. First, 
a larger number of cuts has to be calculated. Second,
in addition to the cut-constructible part we calculate 
also the rational part. Last, the evaluation of tree level amplitudes 
with (massive) quarks takes more computational effort.

\subsection{Scattering amplitudes with two quarks and three gluons}

\begin{tiny}
\begin{table}[th!]
\begin{center}
\begin{tabular}{|c|c|c|c|}
\hline\hline
Amplitude & tree & $c^{\rm cut}$ &  $c$ \\ \hline\hline
$+_{\bar t},+_{t},+_{3},+_{4},+_{5}$ 
&
 -0.000533-0.000137\; i
& 
9.584144+6.530925\; i
& 
51.8809+6.543042\; i
 \\ \hline
$+_{\bar t},-_{t},+_{3},-_{4},+_{5}$ & 
 -0.004540 +  0.018665\; i
&19.65913-11.77003\; i
&
23.00306-9.699584 \; i
\\ \hline
$+_{\bar t},+_{t},-_{3},+_{4},-_{5}$ & 
 -0.004726+  0.014201\; i
&33.15950-1.832717\; i
& 
33.71943 -3.142751\; i
\\ \hline 
$+_{\bar t},-_{t},-_{3},+_{4},+_{5}$ & 
 0.045786  +  0.010661\; i
& 22.84043-6.540697\; i
& 
23.03114-7.313041 \; i
 \\ \hline \hline

$+_{\bar t},+_{3},+_{t},+_{4},+_5$ & 
 0.000182  + 0.001369\; i
&6.517366-1.277070 \;i
&
19.37656+7.563101\;i
\\ \hline
$+_{\bar t},+_{3},-_{t},-_{4},+_5$ & 
 0.0467366-0.006020\; i
&19.440997-7.639466\;i
&  
20.93024-9.936409 \; i
\\ \hline
$+_{\bar t},-_{3},+_{t},+_{4},-_5$ &  
 0.019275 -0.0732138\; i
&15.31910 -3.9278496 \;i 
&  
15.176306-4.102803\;i
\\ \hline
$+_{\bar t},-_{3},-_{t},+_{4},+_5$ & 
 -0.018203-0.111312\; i
& 24.13158+1.431596
\;i
&  
24.70002+1.018096\; i
 \\ \hline \hline

$+_{\bar t},+_{3},+_{4},+_{t},+_5$ & 
   0.00060-0.001377\; i
& 13.13854+6.157043\;i
& 
10.13113+13.83997\;i
  \\ \hline
$+_{\bar t},+_{3},-_{4},-_{t},+_5$ & 
 -0.047199-0.021516\; i
&23.90539 -2.168867\;i
& 
22.905695-4.284617\;i
\\ \hline
$+_{\bar t},-_{3},+_{4},+_{t},-_5$ &

-0.015110+0.063118\; i &
13.54258-7.800591\; i &
13.50273-8.018127\; i
\\ \hline
$+_{\bar t},-_{3},+_{4},-_{t},+_5$ & 
 -0.048800+ 0.112645\; i
& 21.77602+ 2.078051
\;i
& 
22.52784+1.424481\;i
 \\ \hline \hline
$+_{\bar t},+_{3},+_{4},+_5,+_{t}$ & 
  -0.000252+0.000144\; i
& -10.35085+45.26276 \; i
& 
-98.81384+52.81712 \;i
  \\ \hline
$+_{\bar t},+_{3},-_{4},+_5,-_{t}$ & 
  0.0050023+0.008871\; i
& 23.944473+2.862220
\;i
& 
20.92683-0.968026\; i
\\ \hline
$+_{\bar t},-_{3},+_{4},-_5,+_{t}$ &

0.000561-0.004105\; i &
-2.987822-42.00048\; i &
-3.834451-43.67103\; i
\\ \hline
$+_{\bar t},-_{3},+_{4},+_5,-_{t}$ & 
  0.021216-0.011994\; i
& 19.72995-2.120128\; i
& 
20.94996-1.684734\;i 
 \\ \hline \hline
\end{tabular}
\caption{\label{3g} The $0 \mapsto \bar t t\,+\, 3$ gluons results for 
the primitive tree amplitude and finite parts of the four one-loop
primitive amplitudes for various helicities of gluons and top-quarks. 
Both the cut-constructible and total finite one-loop terms are given.}
\vspace{-0.1cm}
\end{center}
\end{table}
\end{tiny}

In Table~\ref{3g} the results for the finite parts of the four primitive
amplitudes
$A_{L}(1_{\bar t},2_{t},g_3,g_4,g_5)$,
$A^{L}(1_{\bar t},g_3,2_{t},g_4,g_5)$, 
$A^{L}(1_{\bar t},g_3,g_4,2_{t},g_5)$ and 
$A^{L}(1_{\bar t},g_3,g_4,g_5,2_{t})$
are presented in the FDH scheme.
We take the mass of the top quark to be $m_t = 1.75$, the scale $\mu=E$ 
and choose the  kinematic point
\ba
&&p_1=E\left ( 1,0,0,\beta \right),\;\;\; p_2= E\left ( 1,0,0,-\beta \right),  \\
&&p_3=E\xi \left ( -1,1,0,0 \right ),\;\;\; 
p_4= E\xi \left (-\sqrt{2}, 0, 1, 1 \right ), \nonumber \\
&&p_5= -p_1 - p_2 - p_3 - p_4, \nonumber
\ea
where $E=10$, $\beta = \sqrt{1-m_t^2/E^2}$ and 
$\xi = 2/(1+\sqrt{2} + \sqrt{3})=0.4823619098$. 

On a standard  
Pentium 2.33~GHz processor it takes about 27, 35, 45 and 50 ms 
respectively to evaluate the primitive amplitudes
$A_L(1_{\bar t}, 2_t, 3,4,5)$, $A_L(1_{\bar t}, 3,2_t,4,5),
A_L(1_{\bar t}, 3,4,2_t,5)$ and $A_L(1_{\bar t}, 3,4,5,2_t)$.
Comparing these evaluation times to the 
$\bar t t\,+\, 2$ gluon evaluation time, we see that the scaling is similar to 
the time scaling of the four and five gluon evaluation time in Ref.~\cite{Ellis:2007br}. Similar to $t \bar t +$ 2 gluons case, 
approximately 
half of the time is spent on the evaluation of the rational part.

\section{Conclusions}

In this paper we extended the method of generalized $D$-dimensional 
unitarity by computing  one-loop scattering amplitudes for 
processes with   massive quarks. We have proposed a solution
to the subtleties associated with
external self-energies and renormalization. We validated the method
by  computing  the one-loop amplitudes  for $0 \mapsto t \bar t\ +$ 2 gluons
and  $0 \mapsto t \bar t\ +$ 3 gluons. We have
shown that the method is amenable to efficient numerical implementation.
The results of this paper show that the 
generalized $D$-dimensional unitarity is a robust computational method.
It will allow us to carry out NLO calculations 
for  a large number of high multiplicity processes 
with massive particles,  relevant for LHC phenomenology.

\section*{Acknowledgments} 
K.M. is supported in part by the DOE grant DE-FG03-94ER-40833.

\end{document}